\documentclass[draft,grl]{agutex}

\usepackage{graphicx}
\setkeys{Gin}{draft=false}
\authorrunninghead{KOSKINEN ET AL.}

\titlerunninghead{CH$_4$ ESCAPE FROM PLUTO}

\authoraddr{Corresponding author: T. T. Koskinen,
Lunar and Planetary Laboratory, University of
Arizona, Kuiper Building 413, Tucson, AZ 85721, USA.
(tommi@lpl.arizona.edu)}

\begin{document}

%
%

\title{On the escape of CH$_4$ from Pluto's atmosphere}
%
%

%
%



\authors{T. T. Koskinen\altaffilmark{1}, J. T. Erwin\altaffilmark{1}, R. V. Yelle\altaffilmark{1}}

\altaffiltext{1}{Lunar and Planetary Laboratory, University of Arizona, 1629 E. University Blvd., Tucson, Arizona, USA.} 

\textbf{Key points:} Escape of CH$_4$ from Pluto is simulated by a multi-species, time-dependent model.  The mixing ratios of CH$_4$ depend strongly on the escape rate.  Observed CH$_4$ profiles constrain the escape rate.

%
%


\begin{abstract}
We adapted a multi-species escape model, developed for close-in extrasolar planets, to calculate the escape rates of CH$_4$ and N$_2$ from Pluto.  In the absence of escape, CH$_4$ should overtake N$_2$ as the dominant species below the exobase.  The CH$_4$ profile depends strongly on the escape rate, however, and the typical escape rates predicted for Pluto lead to a nearly constant mixing ratio of less than 1 \% below the exobase.  In this case the CH$_4$ escape rate is only 5--10 \% of the N$_2$ escape rate.  Observations of the CH$_4$ profile by the New Horizons/ALICE spectrograph can constrain the CH$_4$ escape rate and provide a unique test for escape models. 
\end{abstract}

%
%

%

\begin{article}

%
%

\section{Introduction}
\label{sc:intro}   

Due to its low surface gravity, Pluto has an atmosphere that extends to several radii ($R_p$) and undergoes energy-limited escape.  This means that a significant fraction of the solar EUV radiation absorbed by the upper atmosphere powers mass loss \citep{erwin13,zhu14}.  In this regard Pluto, despite its large distance from the Sun, is similar to many close-in extrasolar planets that undergo rapid, energy-limited escape \citep[e.g.,][]{yelle04,koskinen07,koskinen14}.  The same escape regime may also have shaped the composition of the early atmospheres of Venus, Earth and Mars \citep[e.g.,][]{zahnle86,hunten87,tian08a,tian08b}, and is likely to be important on many potentially habitable rocky exoplanets to be studied in the future \citep[e.g.][]{tarter07,tian09}.  This makes the observations of Pluto's upper atmosphere by the New Horizons/ALICE ultraviolet (UV) spectrograph especially interesting.                   

The escape rate, that is constrained by the observed density and temperature profiles, is particularly important.  It is the most pertinent test on the models, and a critical factor affecting the long-term evolution of the atmosphere.  Most of the recent models, however, focus on N$_2$, the dominant species.  These models show that different escape mechanisms, hydrodynamic or Jeans escape, that are controlled by the upper boundary conditions, lead to different density and temperature profiles that nevertheless correspond roughly to the same mass loss rate \citep[e.g.,][]{erwin13}.  This means that, depending on the model, the same observed density profile can also be matched with different mass loss rates.  Furthermore, the models indicate that the density profiles in the upper atmosphere are sensitive to thermal structure and radiative balance in the lower atmosphere that are not directly related to the escape rate \citep{zhu14}.  These factors complicate the use of the N$_2$ profile to constrain the escape rate.

The ALICE spectrograph covers the wavelength range from 50 nm to 190 nm \citep{stern05}, including the ionization and dissociation continua of N$_2$ and CH$_4$, as well as the N$_2$ electronic bands, that probe the upper atmosphere above $r \approx$~1.5 $R_p$.  Thus the solar occultation measurements that will be obtained during the encounter provide simultaneous coverage of N$_2$ and CH$_4$ at radial distances of about 1.5--3 $R_p$.  The density profile of CH$_4$ is controlled by both diffusion and the escape flux, and it can therefore provide further constraints on the escape rate.  In this work we use a multi-species escape model, developed originally for extrasolar planets \citep{koskinen13a,koskinen13b}, to calculate the escape rates of N$_2$ and CH$_4$ from Pluto at the conditions expected during the New Horizons encounter.

We note that early models of escape from Pluto assumed that CH$_4$ is the dominant species in the upper atmosphere, based on the argument that diffusive separation should lead CH$_4$ to overtake N$_2$ \citep{mcnutt89,hubbard90}.  In contrast, later work by \citet{krasnopolsky99b}, that included the effect of escape on CH$_4$ and several other minor species in addition to diffusion, indicated that N$_2$ is the dominant species at all altitudes.  However, they also showed that the abundance of CH$_4$ is potentially significant near the exobase.  The purpose of this work is to revise the predictions for CH$_4$ escape in light of the recent progress in models of atmospheric escape for Pluto \citep{tucker12,erwin13,zhu14}, and to study the sensitivity of the CH$_4$ density profile to the escape rate, diffusion and surface mixing ratio. 

\section{Methods}
\label{sc:methods}     

A detailed description of our escape model is given in \citet{koskinen13a,koskinen13b}.  The model solves the time-dependent equations of continuity, momentum and energy in the vertical direction.  We note that a time-dependent model passes smoothly through the critical point in transonic escape, although this is not important here because escape from Pluto is subsonic.  We solve common momentum and energy equations for the bulk flow, together with separate continuity equations for the different species i.e., N$_2$, CH$_4$ and CO.  The velocities of the these species are solved self-consistently by using the diffusion approximation \citep[e.g.,][]{garciamunoz07}.  The model is capable of including photochemistry in the upper atmosphere but this option is not used here because its influence on the density profiles should be negligible.

In order to calculate the solar EUV heating rate in the upper atmosphere, we used a Level 3 spectrum of the Sun from the TIMED/SEE archive (http://lasp.colorado.edu/lisird/see/) generated for 28 April, 2015 at a resolution of 1 nm.  We calculated the absorption of radiation self-consistently in the model, and assumed EUV heating efficiencies of 25 \% and 50 \% for N$_2$ and CH$_4$, respectively \citep{krasnopolsky99a}.  This calculation differs from \citet{zhu14} in that we used the full solar spectrum to calculate the EUV heating rate instead of only 3 broad wavelength bins.  We also used the parameterization given by \citet{strobel08} to include infrared (IR) cooling from rotational lines of CO.  We did not include heating and cooling by the near-IR bands of CH$_4$ that should be negligible in the upper atmosphere above our lower boundary at $p_0 =$~10$^{-7}$ bar.   

Boundary conditions are of critical importance to the results.  At the lower boundary we set the temperature to $T_0 =$~103.7 K and altitude to $z_0 =$~340.5 km, in agreement with the reference model of \citet{zhu14}.  For the composition, we varied the volume mixing ratio of CH$_4$ by adopting surface values of 0.3 \%, 0.44 \% and 0.6 \% for different simulations, based on the range derived from VLT/CRIRES observations \citep{lellouch15}.  In line with \citet{zhu14}, we assumed a surface mixing ratio of 5~$\times$~10$^{-4}$ for CO in all of our simulations.  The eddy mixing rate is currently unknown, and we adopted values of 10, 100, and 1000 m$^2$~s$^{-1}$ for the eddy diffusion coefficient $K_{zz}$.  The molecular diffusion coefficients were taken from the compilation of \citet{marrero72}.  Our reference model assumes a CH$_4$ surface mixing ratio of 0.44 \% and $K_{zz} =$~100 m$^2$~s$^{-1}$.  The CH$_4$-N$_2$ diffusion coefficient in the reference model varies from 34 m$^2$~s$^{-1}$ at the lower boundary to 8.5~$\times$~10$^7$ m$^2$~s$^{-1}$ at the exobase.

We can use the above information to verify that the CH$_4$ profile is not significantly affected by photochemistry.  The photolysis of CH$_4$ is dominated by the solar Lyman $\alpha$ photon flux, which is 3.8~$\times$~10$^{12}$ m$^{-2}$~s$^{-1}$ at 32.9 AU based on the TIMED/SEE spectrum.  The CH$_4$ cross section at Lyman $\alpha$ is 1.85~$\times$~10$^{-21}$ m$^2$ and the subsolar optical depth based on our model reaches unity near 1.4 $R_p$.  The subsolar (global minimum) photolysis timescale at 1.4 $R_p$ is therefore 3.9~$\times$~10$^8$ s.  The CH$_4$--N$_2$ diffusion coefficient at that level is $D_{st} =$~1.7~$\times$~10$^2$ m$^2$~s$^{-1}$ and the scale height is 92 km.  Thus the subsolar photolysis timescale is longer than the diffusion timescale of $\tau_d = H^2/D_{st} =$~5~$\times$~10$^7$ s.  This difference increases rapidly with altitude and near the effective heating peak of about 1.8 $R_p$ (see Section~\ref{subsc:mloss}) the photolysis timescale is 50 times longer than the diffusion timescale.             

At the upper boundary, we used the modified Jeans (Type-I) boundary conditions with an enhancement factor of 1.5 \citep{zhu14}.  The upper boundary is placed at the exobase (where the Knudsen number $Kn =$~1), the altitude of which is automatically adjusted by our model while the simulations approach steady state.  We used the same boundary condition for the temperature gradient as \citet{zhu14}.  Their model, however, did not consider CO and CH$_4$ separately whereas our model requires additional boundary conditions for the velocities and partial pressure gradients of the individual species.  

We set the velocities of the different species at the exobase to the Jeans effusion velocity:
\begin{equation}
w_{sJ} = \Gamma \sqrt{ \frac{kT}{2 \pi m_s} } \left( 1 + \lambda_s \right) \exp \left( -\lambda_s \right)
\end{equation}  
where $\lambda_s$ is the escape parameter for a species with mass $m_s$ and $\Gamma =$~1.5.  The pressure gradient for species $s$ at the upper boundary can then be derived from the diffusion equation \citep[e.g.,][]{schunk00}, and it is given by:  
\begin{equation}
\frac{1}{p_s} \frac{\partial p_s}{\partial r} = \frac{m_s}{m} \frac{1}{p} \frac{\partial p}{\partial r} - w_{sJ} \sum_{t \ne s} \frac{x_t}{D_{st}} + \sum_{t \ne s} x_t \frac{w_{tJ}}{D_{st}}
\label{eqn:denb}
\end{equation}
where $p = \sum_s p_s$, $m$ is the mean molecular weight, $x_t$ is the volume mixing ratio and $D_{st}$ is the mutual diffusion coefficient for species $s$ and $t$.  We note that eddy diffusion does not need to be included in the upper boundary conditions because $K_{zz} << D_s$ at the exobase.

\section{Results}
\label{sc:results}

\subsection{CH$_4$ mixing ratios}

\begin{figure}
\noindent\includegraphics[width=40pc]{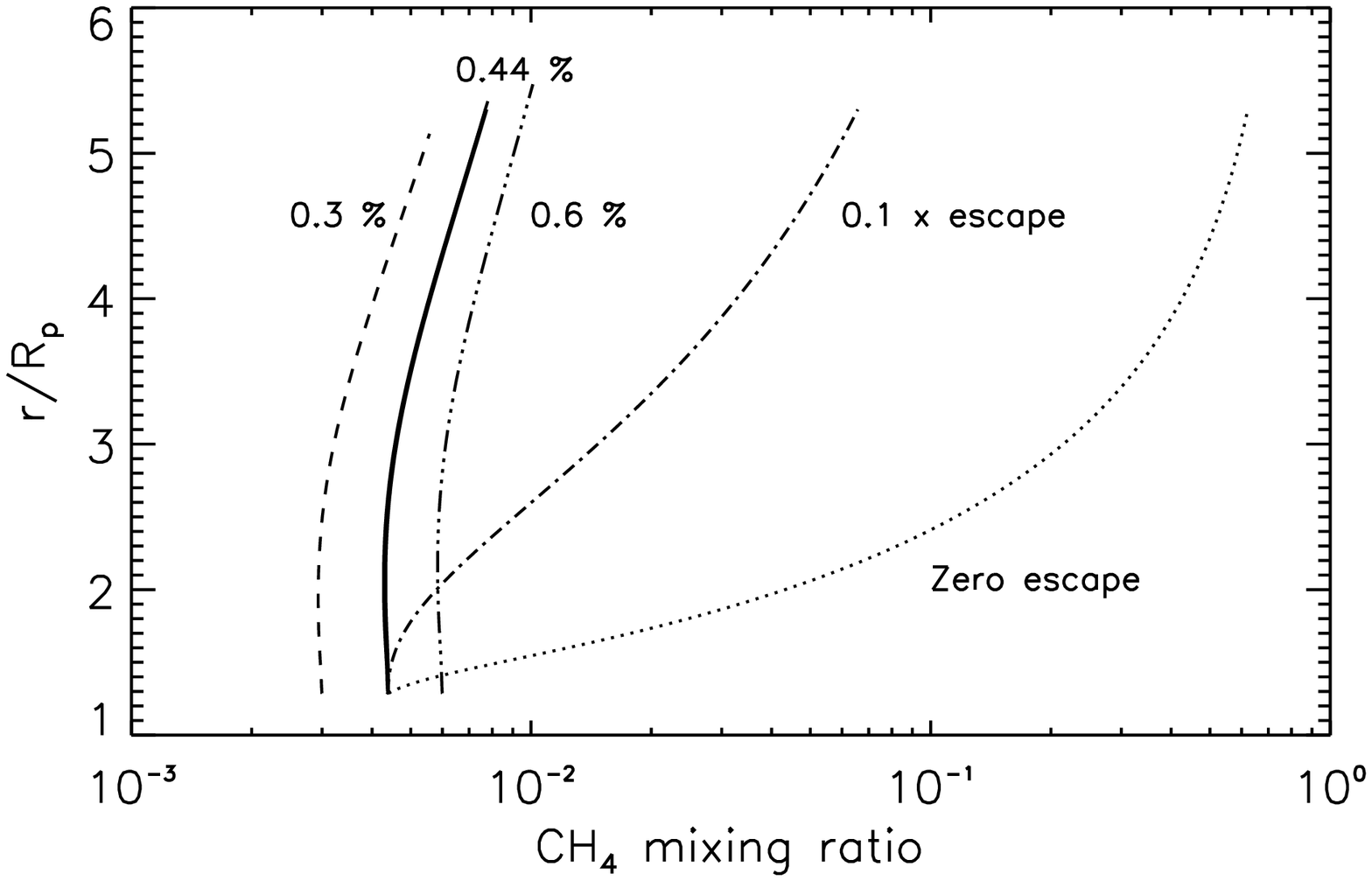}
\caption{The mixing ratio of CH$_4$ in different simulations.  The solid lines (that are indistinguishable) are based on a surface mixing ratio of 0.44 \% and $K_{zz} =$~10--1000 m$^2$~s$^{-1}$.  The dashed and dash-triple-dotted lines are based on surface mixing ratios of 0.3 \% and 0.6 \%, respectively, with $K_{zz} =$~100 m$^2$~s$^{-1}$.  The dotted and dash-dotted lines show the zero escape and reduced escape rate profiles, respectively, based on our reference model temperature-pressure profile, a surface mixing ratio of 0.44 \% and $K_{zz} =$~100 m$^2$~s$^{-1}$.  In each case, the upper boundary is the exobase.}
\label{fig:ch4profs}
\end{figure}

The CH$_4$ abundance profile in Pluto's atmosphere is controlled primarily by escape, with limited sensitivity to our range of surface mixing ratios or $K_{zz}$.  This is illustrated by Figure~\ref{fig:ch4profs} that shows the CH$_4$ profiles for 5 different simulations.  The solid lines show results for models with $K_{zz} =$~10--1000 m$^2$~s$^{-1}$ and a surface mixing ratio of 0.44 \%, including the reference model, while the dashed and dashed-triple-dotted lines show models with $K_{zz} =$~100 m$^2$~s$^{-1}$ and CH$_4$ surface mixing ratios of 0.3 \% and 0.6 \%, respectively.  The global escape rate in these models, which increases with the surface mixing ratio of CH$_4$, is 1.8--2.3~$\times$~10$^{27}$ s$^{-1}$.  This is the sum of the escape rates of the individual species (hereafter, the total escape rate).  For comparison, the dotted and dash-dotted lines show the zero escape and reduced escape CH$_4$ profiles, respectively, based on the T-P profile in the reference model.  The reduced total escape rate here is  2.9~$\times$~10$^{26}$ s$^{-1}$, based on a reduced Jeans enhancement factor of $\Gamma =$~0.15.       

In the absence of escape CH$_4$ would become the dominant species in the reference model around $r =$~4.4 $R_p$ i.e., well below the exobase, which is located at $r =$~5--6 $R_p$.  With escape included, however, the CH$_4$ mixing ratio is less than 1 \% at all altitudes.  Near $r =$~3 $R_p$ where solar occultations in the EUV become sensitive to absorption by CH$_4$ (see Section~\ref{subsc:obs}), the mixing ratio in the reference model is only 0.46 \% whereas in the zero escape case it is almost 50 times larger at 21.4 \%.  An escape rate of about 2~$\times$~10$^{27}$ s$^{-1}$ therefore leads to a nearly constant mixing ratio of CH$_4$ below $r =$~3 $R_p$.  

The strong dependence of the CH$_4$ mixing ratios on escape means that observations of the CH$_4$ profile provide a means to constrain the escape rate and thus the first opportunity to partially validate a hydrodynamic escape model for a planetary atmosphere.  The escape rate predicted by the current models is near the saturation value i.e., the value that leads to a nearly constant mixing ratio, which is close to the surface mixing ratio, below 3 $R_p$.  As shown by Figure~\ref{fig:ch4profs}, a lower total escape rate of 2.9~$\times$~10$^{26}$ s$^{-1}$ with the reference model T-P profile leads to a higher CH$_4$ mixing ratio of 1.5 \% at $r =$~3 $R_p$.  We note that the CH$_4$ profiles in Figure~\ref{fig:ch4profs} are practically independent of $K_{zz}$, which further enhances their potential to constrain the escape rate.       

We note that our results differ from \citet{krasnopolsky99b} who also studied the diffusion and escape of CH$_4$ on Pluto.  Their model, which included an N$_2$ escape rate of 2.6~$\times$~10$^{27}$ s$^{-1}$, predicted that diffusive separation of CH$_4$ takes place near $r =$~2300 km, and that the mixing ratio of CH$_4$ is about 19 \% at $r =$~3500 km.  In our models the mixing ratio of CH$_4$ at $r =$~3500 km varies between 0.32 \% and 0.61 \%.  The differences between our model and the earlier work are difficult to identify exactly.  The exobase, however, was typically at a lower altitude in the model of \citet{krasnopolsky99b}, and their temperatures were generally cooler.  Our temperatures are significantly warmer due to the upper boundary condition based on the Jeans energy flux.                                                    

\subsection{Temperature profiles}

\begin{figure}
\noindent\includegraphics[width=40pc]{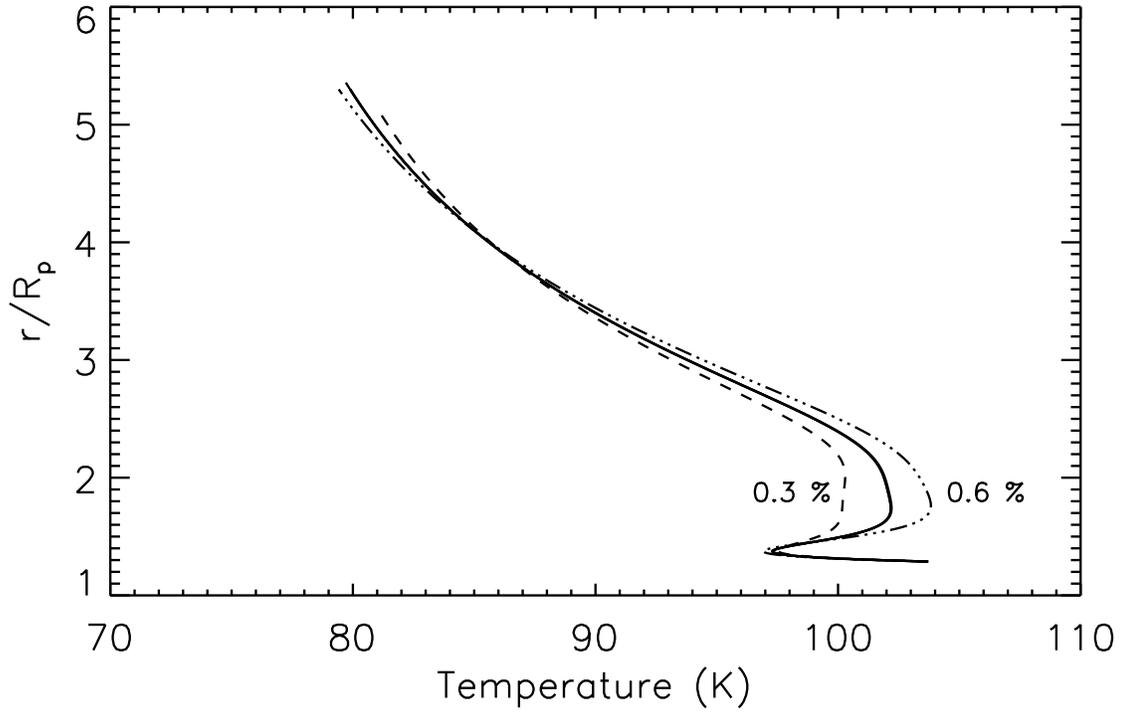}
\caption{Temperature profiles based on simulations with different surface mixing ratios and $K_{zz}$, corresponding to the CH$_4$ mixing ratio profiles in Figure~\ref{fig:ch4profs}.  The solid lines (that are indistinguishable) show results for models with a CH$_4$ surface mixing ratio 0.44 \% and $K_{zz} =$~10--1000 m$^2$~s$^{-1}$.  The dashed lines and dash-triple-dotted lines show results based on CH$_4$ surface mixing ratios of 0.3 \% and 0.6 \%, respectively, with $K_{zz} =$~100 m$^2$~s$^{-1}$.}
\label{fig:temps}
\end{figure}

The temperature profiles based on our models are shown by Figure~\ref{fig:temps}.  The differences in the predicted temperatures are within 5 K.  The peak temperature increases with the abundance of CH$_4$, while the temperature near the exobase decreases slightly as the peak temperature increases.  This is caused by more efficient absorption of Lyman $\alpha$ radiation by CH$_4$, which leads to both higher peak temperatures and a slightly higher escape rate.  As a result, adiabatic cooling near the exobase is also more efficient, leading to a slightly lower temperature near the exobase.  Due to the nearly constant mixing ratio of less than 1 \% for CH$_4$ in the model, however, the inclusion of diffusion does not significantly affect the results of \citet{zhu14} who assumed that CH$_4$ is uniformly mixed with the surface value of 0.44 \%.  Our temperature profiles do differ slightly from their predictions, but this is mostly due to the CO cooling parameterization in our work that leads to lower peak temperatures.                  

\subsection{Energy balance and CH$_4$ escape}
\label{subsc:mloss}

Our reference model predicts a total escape rate of 2.1~$\times$~10$^{27}$ s$^{-1}$.  This value is essentially identical to the previous estimates of the escape rate from Pluto's atmosphere.  We find that N$_2$ and CO escape from the exobase with an outflow velocity of 1.75 m~s$^{-1}$, while the CH$_4$ outflow velocity is much higher, about 18 m~s$^{-1}$.  Due to the relatively low abundance of CH$_4$, however, the effective bulk outflow velocity i.e., $w = \sum_s \rho_s w_s/\rho$, is still only 1.82 m~s$^{-1}$.  The predicted escape rate of CH$_4$ in our models is 1--2~$\times$~10$^{26}$ s$^{-1}$, or about 5--10 \% of the N$_2$ escape rate.  These values are an order of magnitude lower than the diffusion limit of 2~$\times$~10$^{27}$ s$^{-1}$ \citep{strobel08}.  Thus our reference model agrees with recent work in that escape from Pluto is strongly subsonic, and that the escape of N$_2$ dominates \citep[e.g.,][]{erwin13,zhu14}.  

These models also show that mass loss from Pluto is energy-limited.  This statement needs to be qualified, because of the efficiency factors that affect this conclusion.  The energy limited mass loss rate based on globally averaged solar flux is:
\begin{equation}
\dot{M} = \frac{\eta_E \alpha_E \pi r_E^2 F_E}{\Delta \Phi}
\end{equation}              
where $F_E$ is the integrated solar flux, in this case at wavelengths of 1--145 nm, $\eta_E$ is the mass loss efficiency, $\alpha_E$ is the heating efficiency, $r_E$ is the effective radius of the heating peak and $\Delta \Phi$ is the gravitational potential difference that the escaping particles have to overcome.  

At the distance of 32.9 AU, the integrated solar flux is $F_E =$~1.18~$\times$~10$^{-5}$ W~m$^{-2}$, based on the TIMED/SEE spectrum.  The global, effective heat flux in our reference model is $\alpha_E F_E =$~3.01~$\times$~10$^{-6}$ W~m$^{-2}$, which is based on the net heating rate that includes CO cooling, and $r_E =$~1.78 $R_p$.  Thus the heating efficiency is $\alpha_E \approx$~0.26, and the energy-limited mass loss rate based on this efficiency would be 103 kg~s$^{-1}$.  The mass loss rate predicted by the reference model, on the other hand, is 96 kg~s$^{-1}$, implying that the mass loss efficiency is $\eta_E \approx$~0.93.  This means that, at least according to the models, escape from Pluto is energy-limited i.e., $\eta_E \rightarrow$~1, in analogy to many close-in extrasolar giant planets \citep[e.g.,][]{koskinen14}, and in contrast to the present planetary atmospheres in the solar system.                

\subsection{Boundary conditions}      

Boundary conditions for escape models have been discussed extensively in recent literature on Pluto, but so far this discussion has received only cursory attention elsewhere.  For example, many of the existing time-dependent exoplanet models rely on the so-called outflow boundary conditions i.e., extrapolation of the density, temperature and velocity with a constant slope at the upper boundary \citep[e.g.,][]{tian05,garciamunoz07}.  These boundary conditions lead to unphysical outcomes on Pluto, which has an exobase well below the altitude of the implied sonic point.  In theory, the outflow boundary conditions may be valid above the sonic point, although this assertion probably deserves further attention in future studies.        

Other exoplanet models have imposed Jeans or modified Jeans outflow velocities at the upper boundary with extrapolated temperatures, either with a zero gradient or a constant slope \citep[e.g.,][]{tian08a,tian08b,koskinen14}.  This approach appears to be valid only if the atmosphere is rendered isothermal by conduction at the exobase, in which case the constant slope also reduces to the isothermal boundary condition and produces acceptable results.  On Pluto, extrapolating the temperature with a constant slope leads to an almost adiabatic decrease in temperature with altitude, and a much cooler exobase.  As a result, the conductive heat flux becomes larger than the Jeans energy flux.  Zero gradient, on the other hand, leads to an overestimated temperature near the exobase.  In general, the validity of the boundary conditions in the present models, and thus the escape solution, can be tested by matching the models with both the N$_2$ and CH$_4$ density profiles that are, at least in theory, retrievable from the EUV solar occultations.     

\subsection{Observations}  
\label{subsc:obs} 

The model calculations show that the CH$_4$ profile can be used as an indicator of the escape rate; however, it is important to have coincident measurements of N$_2$ because this enables calculation of the CH$_4$ mixing ratio and the atmospheric temperature.  We note that stellar and solar occultations observed by the Cassini/UVIS instrument in its EUV channel have been used to retrieve coincident density profiles of N$_2$ and CH$_4$ in Titan's upper atmosphere for the same range of slant column densities that is relevant for escape models on Pluto.  In particular, \citet{capalbo13} used the 58.4 nm and 63 nm solar emission lines in the ionization continuum of N$_2$, together with a wavelength bin around 108.5 nm in the ionization continuum of CH$_4$, to retrieve these density profiles from a solar occultation.  In addition, \citet{kammer13} used EUV stellar occultations at 91.1--110 nm, probing the N$_2$ electronic band system, for the same purpose.  

\begin{figure}
\noindent\includegraphics[width=40pc]{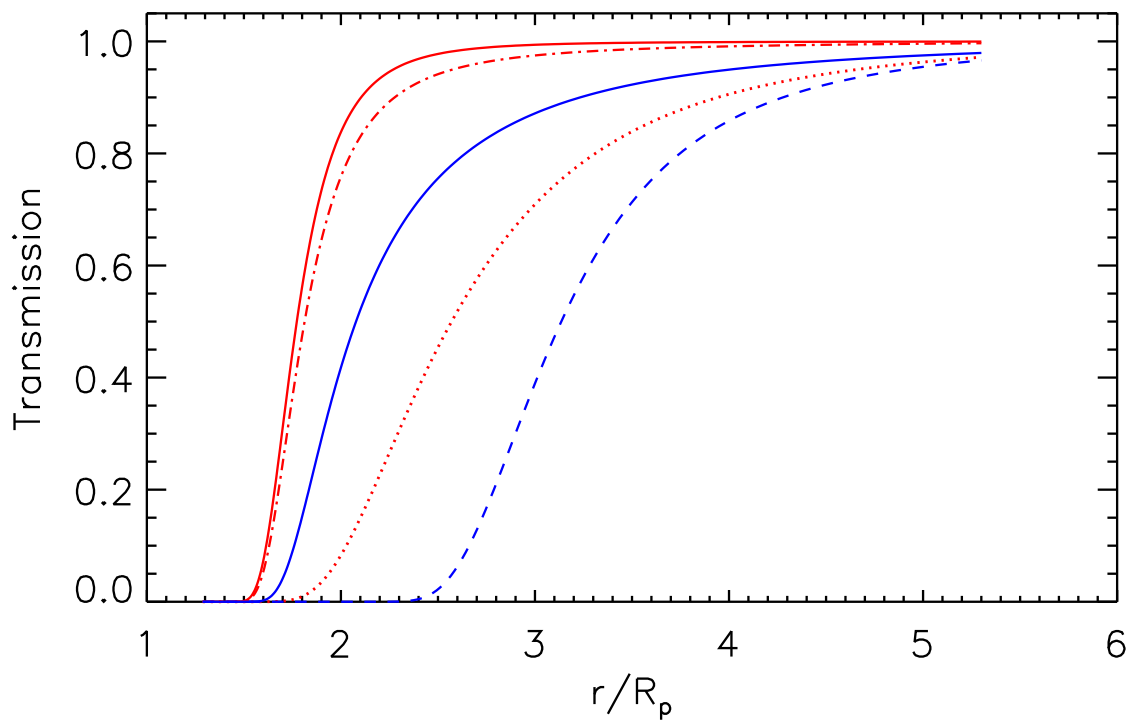}
\caption{Predicted transmission for different EUV wavelength bins.  The red lines show transmission in the 108.5 (108--109) nm bin based on the CH$_4$ profiles in the reference (solid line), zero escape (dotted line) and reduced escape (dash-dotted line) models (see Figure~\ref{fig:ch4profs}).  The blue lines show transmission in the 63 (62.5--63.5) nm bin (dashed line) and 90 (89--90.5) nm bin (solid line) based on the reference model N$_2$ and CH$_4$ profiles.}
\label{fig:lcs}
\end{figure}

For example, we used our models to calculate predicted transmission for Pluto in the 108.5 nm bin, which includes a group of solar N II and He II emission lines \citep{curdt01}.  In this region absorption is by CH$_4$ and the cross section is roughly constant with wavelength \citep{kameta02}, allowing for a simple transmission calculation instead of a more complex forward model.  The results are shown in Figure~\ref{fig:lcs}, which indicates that the zero escape case is clearly distinguishable from our reference model.  The difference in transmission between the reference model and the reduced escape model is less pronounced, peaking at about 0.09 around 1.9 $R_p$ where transmission is about 0.7.  This difference is detectable with a signal-to-noise ratio (S/N) of 8 or higher.     

We also used our reference model to calculate transmission in the 63 nm wavelength bin, which includes the solar O V line that is absorbed predominantly by N$_2$, with a small contribution from CH$_4$.  The resulting light curve is limited to altitudes above 2.5 $R_p$ (Figure~\ref{fig:lcs}).  At lower altitudes the N$_2$ electronic band system (80--100 nm) can be used to retrieve the N$_2$ profile.  To illustrate this, we calculated transmission at 89--90.5 nm in the solar Lyman continuum, in order to avoid uncertainties associated with the width of the solar emission lines.  Some of these lines are absorbed by the N$_2$ bands while absorption by CH$_4$ features strongly between the bands.  A high resolution forward model is therefore required to calculate transmission, even when instrument broadening is significant.  In our model we used a SUMER spectrum of the Sun \citep{curdt01} and a high resolution N$_2$ band cross section \citep{lewis08}.  In order to include the point spread function (PSF) for ALICE, we used a Gaussian with a FWHM of 0.4 nm \citep{stern05}.  The resulting occultation light curves in Figure~\ref{fig:lcs} show that CH$_4$ can be measured from 1.6 $R_p$ to 2--4 $R_p$ and N$_2$ can be measured from 1.8 $R_p$ to 4 $R_p$, assuming that densities can be retrieved for transmissions between 0.1 and 0.9.  The precise altitude range will, of course, depend on the noise characteristics of the data set.              

\section{Conclusions}
\label{sc:discussion} 

We adapted a multi-species model of hydrodynamic escape to calculate the escape rates of N$_2$ and CH$_4$ from Pluto, and to quantify the effect of escape on the CH$_4$ profile.  The CH$_4$ profile plays a critical role in the heating and photochemistry of the upper atmosphere, and under diffusive separation it should overtake N$_2$ as the dominant species below the exobase.  As we showed, it can also be used to constrain the escape rate.  This provides a unique opportunity to test models of hydrodynamic escape in an actual atmosphere.

The total escape rate predicted by our reference model is 2.1~$\times$~10$^{27}$ s$^{-1}$, in agreement with previous models.  In terms of energy-limited escape, this corresponds to $\sim$24 \% of the incident solar UV flux at 1--145 nm with an effective heating peak around 1.78 $R_p$.  The CH$_4$ escape rate in the model is about 5--10 \% of the N$_2$ escape rate.  We find that the mixing ratio of CH$_4$ in the upper atmosphere depends strongly on the escape rate and is almost independent of $K_{zz}$.  The total escape rate of 2.1~$\times$~10$^{27}$ s$^{-1}$ leads to a nearly constant mixing ratio that is close to the surface value below $r =$~3 $R_p$ whereas lower escape rates lead to substantially higher mixing ratios.  Given an independent measurement of the surface mixing ratio, the CH$_4$ and N$_2$ profiles at $r=$~1.8--3 $R_p$ that are retrievable from EUV occultations can therefore constrain the escape rate.    


%
%
%
%
%
%
%

\begin{acknowledgments}
The authors would like to thank the anonymous reviewers for their insightful comments and suggestions that have contributed to improve this paper.  No data was used in producing this manuscript.  We also thank M. J. Harris for useful correspondence and support with the escape model.
\end{acknowledgments}

%
%
%
%
%
%
%
%
%

\clearpage



%

%
%
\end{article}
%
%
%
%
%
%

%



\begin{thebibliography}{}

\providecommand{\natexlab}[1]{#1}
\expandafter\ifx\csname urlstyle\endcsname\relax
  \providecommand{\doi}[1]{doi:\discretionary{}{}{}#1}\else
  \providecommand{\doi}{doi:\discretionary{}{}{}\begingroup
  \urlstyle{rm}\Url}\fi

\bibitem[{\textit{Capalbo et~al.}(2013)}]{capalbo13} Capalbo, F. J., et al. (2013),  Solar occultation by Titan measured by Cassini/UVIS, \textit{Astrophys. J. Lett.}, \textit{766}, L16.
\bibitem[{\textit{Curdt et~al.}(2001)}]{curdt01} Curdt, W., et al. (2001), The SUMER spectral atlas of solar-disk features, \textit{Astron. Astrophys.}, \textit{375}, 591--613.
\bibitem[{\textit{Erwin et~al.}(2013)}]{erwin13} Erwin, J., O.~J. Tucker, and R.~E. Johnson (2013), Hybrid fluid/kinetic modeling of Pluto's escaping atmosphere, \textit{Icarus}, \textit{226}, 375--384.  
\bibitem[{\textit{Garc\'ia Mu\~noz}(2007)}]{garciamunoz07} Garc\'ia Mu\~noz, A. (2007), Physical and chemical aeronomy of HD209458b, \textit{Plan. Space Sci.}, \textit{55}, 1426--1455.
\bibitem[{\textit{Hubbard et~al.}(1990)}]{hubbard90} Hubbard, W.~B., R.~V. Yelle, and J.~I. Lunine (1990), Nonisothermal Pluto atmosphere models, \textit{Icarus}, \textit{84}, 1--11.
\bibitem[{\textit{Hunten et~al.}(1987)}]{hunten87} Hunten, D.~M., R.~O. Pepin, and J.~C.~G. Walker (1987), Mass fractionation in hydrodynamic escape, \textit{Icarus}, \textit{69}, 532--549.
\bibitem[{\textit{Kameta et~al.}(2002)}]{kameta02} Kameta, K., N. Kouchi, M. Ukai, and Y. Hatano (2002), Photoabsorption, photoionization, and neutral-dissociation cross sections of simple hydrocarbons in the vacuum ultraviolet range, \textit{J. Electr. Spectr. Rel. Phen.}, \textit{123}, 225--238.
\bibitem[{\textit{Kammer et~al.}(2013)}]{kammer13} Kammer, J. A., D.~E. Shemansky, X. Zhang, and Y.~L. Yung (2013), Composition of Titan's upper atmosphere from Cassini UVIS EUV stellar occultations.  \textit{Plan. Space. Sci.}, \textit{88}, 86--92.
\bibitem[{\textit{Koskinen et~al.}(2007)}]{koskinen07} Koskinen, T. T., A.~D. Aylward, and S. Miller (2007), A stability limit for the atmospheres of giant extrasolar planets, \textit{Nature}, \textit{450}, 845--848.
\bibitem[{\textit{Koskinen et~al.}(2013a)}]{koskinen13a} Koskinen, T. T., M.~J. Harris, R.~V. Yelle, and P. Lavvas (2013a), The escape of heavy atoms from the ionosphere of HD209458b. I. A photochemical-dynamical model of the thermosphere, \textit{Icarus}, \textit{226}, 1678--1694.
\bibitem[{\textit{Koskinen et~al.}(2013b)}]{koskinen13b} Koskinen, T. T., R.~V. Yelle, M.~J. Harris, and P. Lavvas (2013b), The escape of heavy atoms from the ionosphere of HD209458b. II. Interpretation of the observations,  \textit{Icarus}, \textit{226}, 1695--1708.
\bibitem[{\textit{Koskinen et~al.}(2014)}]{koskinen14} Koskinen, T. T., P. Lavvas, M.~J. Harris, and R.~V. Yelle (2014), Thermal escape from extrasolar giant planets, \textit{Phil. Trans. R. Soc. A}, \textit{372}, 20130089.
\bibitem[{\textit{Krasnopolsky}(1999)}]{krasnopolsky99a} Krasnopolsky, V. A. (1999), Hydrodynamic flow of N$_2$ from Pluto, \textit{J. Geophys. Res.}, \textit{104}, 5955--5962.
\bibitem[{\textit{Krasnopolsky and Cruikshank}(1999)}]{krasnopolsky99b} Krasnopolsky, V.~A., and D.~P. Cruikshank (1999), Photochemistry of Pluto's atmosphere and ionosphere near perihelion, \textit{J. Geophys. Res.}, \textit{104}, 21979--21996.
\bibitem[{\textit{Lellouch et~al.}(2015)}]{lellouch15} Lellouch, E., et al. (2015), Exploring the spatial, temporal, and vertical distribution of methane in Pluto's atmosphere, \textit{Icarus}, \textit{246}, 268--278.
\bibitem[{\textit{Lewis et~al.}(2008)}]{lewis08} Lewis, B.~R., A.~N. Heays, S.~T. Gibson, H. Lefebvre-Brion, and R. Lefebvre (2008), A coupled-channel model of the $^3 \Pi_u$ states of N$_2$: Structure and interactions of the $3s \sigma_g F_3 \ ^3 \Pi_u$ and $3p \pi_u G_3 \ ^3 \Pi_u$ Rydberg states, \textit{J. Chem. Phys.}, \textit{129}, 164306.
\bibitem[{\textit{Marrero and Mason}(1972)}]{marrero72} Marrero, T.~R., and E.~A. Mason (1972), Gaseous diffusion coefficients, \textit{J. Phys. Chem. Ref. Data}, \textit{1}, 3--118.
\bibitem[{\textit{McNutt}(1989)}]{mcnutt89} McNutt, R.~L.~Jr. (1989), Models of Pluto's upper atmosphere, \textit{Geophys. Res. Lett.}, \textit{16}, 1225--1228.
\bibitem[{\textit{Schunk and Nagy}(2000)}]{schunk00} Schunk, R.~W., and A.~F. Nagy (2000), \textit{Ionospheres: Physics, Plasma Physics, and Chemistry}, Cambridge University Press, Cambridge, England.
\bibitem[{\textit{Stern et~al.}(2005)}]{stern05} Stern, S. A., et al. (2005), Alice: The ultraviolet imaging spectrograph aboard the New Horizons Pluto mission spacecraft, \textit{SPIE}, \textit{5906}, 358--367.
\bibitem[{\textit{Strobel}(2008)}]{strobel08} Strobel, D. F. (2008), N$_2$ escape rates from Pluto's atmosphere, \textit{Icarus}, \textit{193}, 612--619.
\bibitem[{\textit{Tarter et~al.}(2007)}]{tarter07} Tarter, J. C., et al. (2007), A reappraisal of the habitability of planets around M dwarf stars, \textit{Astrobiology}, \textit{7}, 30--65.
\bibitem[{\textit{Tian et al.}(2005)}]{tian05} Tian, F., O.~B. Toon, A.~A. Pavlov, and H. de Sterck (2005), Transonic hydrodynamic escape of hydrogen from extrasolar planetary atmospheres, \textit{Astrophys. J.}, \textit{621}, 1049--1060.
\bibitem[{\textit{Tian et al.}(2008a)}]{tian08a} Tian, F., J.~F. Kasting, H.-L. Liu, and R.~G. Roble (2008a), Hydrodynamic planetary thermosphere model: 1. Response of the Earth's thermosphere to extreme solar EUV conditions and the significance of adiabatic cooling, \textit{J. Geophys. Res.}, \textit{113}, E05008.
\bibitem[{\textit{Tian et al.}(2008b)}]{tian08b} Tian, F., S.~C. Solomon, L. Qian, J. Lei, and R.~G. Roble (2008b), Hydrodynamic planetary thermosphere model: 1. Coupling of an electron transport/energy deposition model, \textit{J. Geophys. Res.}, \textit{113}, E07005.
\bibitem[{\textit{Tian}(2009)}]{tian09} Tian, F. (2009), Thermal escape from super Earth atmospheres in the habitable zones of M stars, \textit{Astrophys. J.}, \textit{703}, 905--909.
\bibitem[{\textit{Tucker et al.}(2012)}]{tucker12} Tucker, O. J., J.~T. Erwin, J.~I. Deighan, A.~N. Volkov, and R.~E. Johnson (2012), Thermally driven escape from Pluto's atmosphere: A combined fluid/kinetic model, \textit{Icarus}, \textit{217}, 408--415.
\bibitem[{\textit{Yelle}(2004)}]{yelle04} Yelle, R. V. (2004), Aeronomy of extra-solar giant planets at small orbital distances, \textit{Icarus}, \textit{170}, 167--179.
\bibitem[{\textit{Zahnle and Kasting}(1986)}]{zahnle86} Zahnle, K.~J., and J.~F. Kasting (1986), Mass fractionation during transonic escape and its implications for loss of water from Mars and Venus, \textit{Icarus}, \textit{68}, 462--480. 
\bibitem[{\textit{Zhu et al.}(2014)}]{zhu14} Zhu, X., D.~F. Strobel, and J. T. Erwin (2014), The density and thermal structure of Pluto's atmosphere and associated escape processes and rates, \textit{Icarus}, \textit{228}, 301--314.
\end{thebibliography}
\end{document}